\documentclass[12pt,onecolumn]{emulateapj}
\usepackage{rotating}
\usepackage{amsmath}
\usepackage{amssymb}
\usepackage{amsfonts}
\usepackage{epsfig}
\usepackage{natbib}
\usepackage{pdfpages}
\input epsf
\def\1{\'\i}

\begin{document}
\title{Constraining the redshift evolution of the Cosmic Microwave Background
black-body temperature with PLANCK data.} 
\author{
I. de Martino\altaffilmark{1}, R. G\'enova-Santos\altaffilmark{2}, 
F. Atrio-Barandela\altaffilmark{1}, H. Ebeling\altaffilmark{3}, 
A. Kashlinsky\altaffilmark{4}, D. Kocevski\altaffilmark{5}, 
C.J.A.P. Martins\altaffilmark{6,7}
}

\altaffiltext{1}{F\'{\i}sica Te\'orica, Universidad de Salamanca, 37008 Salamanca, Spain;
email: ivan.demartino1983@gmail.com; atrio@usal.es}
\altaffiltext{2}{Instituto de Astrof\1sica de Canarias, 38200 La Laguna, Spain;
email:ricardo.genova@iac.es}
\altaffiltext{3}{ Institute for Astronomy, University of Hawaii, 
Honolulu, HI 96822 USA; email:ebeling@ifa.hawaii.edu}
\altaffiltext{4}{NASA Goddard Space Flight Center and SSAI, Observational Cosmology Lab, 
Greenbelt, MD 20771 USA; email:Alexander.Kashlinsky@nasa.gov}
\altaffiltext{5}{Chemistry-Physics Building, University of Kentucky,
Lexington, KY 40508 USA; email:kocevski@pa.uky.edu}
\altaffiltext{6}{Centro de Astrof\'{\i}sica da Universidade do Porto,
Rua das Estrelas, 4150-762 Porto, Portugal; email: Carlos.Martins@astro.up.pt}
\altaffiltext{7}{Instituto de Astrof\'{\i}sica e Ci\^encias do Espa\c co,
CAUP, Rua das Estrelas, 4150-762 Porto, Portugal}

\begin{abstract}
We constrain the deviation of adiabatic evolution of the Universe
using the data on the Cosmic Microwave Background (CMB) temperature
anisotropies measured by the {\it Planck} satellite and a sample 
of 481 X-ray selected clusters with spectroscopically measured redshifts.
To avoid antenna beam effects, we bring all the maps to the same resolution.
We use a CMB template to subtract the cosmological signal while preserving
the Thermal Sunyaev-Zeldovich (TSZ) anisotropies; next, we remove
galactic foreground emissions around each cluster and we mask out 
all known point sources. If the CMB black-body temperature scales 
with redshift as $T(z)=T_0(1+z)^{1-\alpha}$, we constrain deviations of
adiabatic evolution to be $\alpha=-0.007\pm 0.013$, consistent with the 
temperature-redshift relation of the standard cosmological model.
This result could suffer from a potential bias $\delta\alpha$  associated with 
the CMB template, that we quantify it to be $|\delta\alpha|\le 0.02$
and with the same sign than the measured value of $\alpha$, but is
free from those biases associated with using TSZ selected clusters; 
it represents the best constraint to date of the temperature-redshift 
relation of the Big-Bang model using only CMB data, 
confirming previous results. 
\end{abstract}


\maketitle

\section{Introduction}

Adiabatic expansion and photon number conservation have
produced a Cosmic Microwave Background (CMB) with a black-body 
temperature of $T_0=2.725\pm 0.002$K \citep{fixsen} that
evolves with redshift $z$ as $T(z)=T_0(1+z)$. This temperature-redshift
relation is an important test of the Big-Bang paradigm \citep{avgoustidis}
and of spatial homogeneity \citep{clarkson}. 
Models like decaying vacuum energy density and gravitational `adiabatic' 
photon creation predict a more general scaling \citep{overduin, matyjasek,puy,jetzer2011}.
An imprint on the $T(z)$ relation can be produced if 
the period of accelerated expansion is driven by a phase transition
\citep{mortonson, nunes}. The non-conservation of the photon number density 
changes the temperature-redshift and the distance duality relations. 
Two functional forms have been considered in the literature
$T(z)=T_0(1+z)^{1-\alpha}$  \citep{lima} and $T(z)=T_0(1+bz)$ \citep{losecco}, 
with $\alpha$ and $b$ constant parameters; at low redshifts, the differences between 
both functional forms are small and only the first scaling is usually tested.
In most scenarios, deviations
of adiabatic evolution are associated with distortions of the CMB black-body 
spectrum \citep{chluba} and are strongly constrained by the current FIRAS upper limit 
of \citet{fixsen}. A departure of the standard temperature-redshift relation
would represent an important challenge to the current cosmological model
and it represents a test of these alternative scenarios. 

The earliest measurements on CMB black-body temperature evolution were
obtained using the relative populations of atomic fine-structure levels 
which are excited by the background radiation \citep{songaila,srianand}. 
\citet{noterdaeme} measured $T(z=2.69)=10.5^{0.8}_{-0.6}$K
using the rotational excitation of CO molecules in quasar spectral lines, value 
compatible with $10.06$K, the black-body temperature expected at that redshift
with the standard temperature-redshift relation. Using quasar spectral lines,
the best constraint on deviations from adiabatic evolution at present is
$\alpha=0.009\pm 0.019$ \citep{muller13}; they also obtained
a very stringent individual measurement of the CMB temperature: 
$T(z=0.89)=5.08 \pm 0.10$K. Recently the Planck Collaboration
obtained a much tighter constraint \citep{planck_15_XIII} 
by including data on large scale structure to the CMB data but their measurement 
$\alpha=(0.2\pm 1.1)\times 10^{-3}$ applies to models where the deviation
from adiabatic evolution starts at the last scattering surface. 
The temperature-redshift relation 
has also been probed using the Thermal Sunyaev-Zeldovich
effect (Sunyaev \& Zeldovich, 1970; hereafter TSZ). The TSZ anisotropy 
induced by clusters of galaxies along the line of sight $\hat{n}$
is $\Delta T(\hat{n})=T_0G(x)Y_c$. The Comptonization
parameter is defined as $Y_c=(k_B/m_ec^2)\int T_e d\tau$, where
$d\tau=\sigma_Tn_edl$ is the cluster optical depth,
$n_e, T_e$ are the electron density and temperature
evaluated along the line of sight $l$, $\sigma_T$ the Thomson cross section,
$k_B$ the Boltzmann constant, $m_e$ the electron mass and $c$ the speed of light.
This effect depends on the frequency of observation $\nu_0$ as
$x=h\nu_0(1+z)/k_BT$, where $h$ is the Planck constant
and $T=T(z)$ the CMB black-body temperature at the cluster
location. In the non-relativistic limit, $G(x)= x{\rm coth}(x/2)-4$.
This spectral dependence is different from that of all known foregrounds, 
making the TSZ effect an effective tool for detecting clusters as well
as a potential probe of the redshift evolution of the background temperature.

If the Universe evolves adiabatically then $x$ is independent of redshift.
We will test deviations from adiabatic evolution assuming the \cite{lima}
parametrization, that is, $x=h\nu_0(1+z)^\alpha/k_BT_0$ and $G=G(\nu,\alpha)$. 
\citet{fabbri} proposed to constrain $\alpha$ by measuring the zero cross frequency 
of clusters at different redshifts that, for adiabatic evolution, occurs at 
$\nu\simeq 217GHz$. The measurement of the cross-over frequency is problematic 
since the TSZ is inherently weak and could be dominated by uncertain systematics.
As an alternative, early studies fit the TSZ signal at different frequencies to measure
the function $G(\nu,\alpha)$. We shall denote this procedure the {\it fit method}.
\citet{rephaeli} suggested the {\it ratio method}, constraining $\alpha$ by using
the ratio of the TSZ anisotropy at 
different frequencies, $R(\nu_1,\nu_2,\alpha)= G(\nu_1,\alpha)/G(\nu_2,\alpha)$. 
By taking ratios, the dependence on the Comptonization parameter 
is removed and there is no need to account for model uncertainties on the gas density
and temperature profiles. On the negative side, the analysis is 
more complicated since the distribution of the temperature anisotropy ratios is 
highly non-gaussian \citep{luzzi} and the Kinematic Sunyaev-Zeldovich 
(Sunyaev \& Zeldovich, 1972; hereafter KSZ) generated by cluster peculiar velocities 
is a component always present at the cluster location that can not be
separated from the cosmological signal. For small number of clusters the 
effect needs to be taken into 
account while it is expected to be less important for large cluster samples.
\citet{battistelli} found $\alpha=-0.16^{+0.34}_{-0.32}$ from the 
TSZ measurement of the Coma and A2163 clusters. Later, from the 
data on just 13 clusters, \citet{luzzi} set up an upper limit of $\alpha\le 0.079$ 
in the range $z=0.023-0.546$  at the 68\% confidence level.

In 2013 the Planck Collaboration released their first maps of the 
temperature fluctuations of the CMB sky using 15.5 months of data 
\citep{planck_01}. The Low Frequency Instrument (LFI) produced three maps
with frequencies 30, 44 and 70~GHz and angular resolution from 33 to
$13\arcmin$. The High Frequency Instrument (HFI) was
sensitive to a wider range of frequencies, from 100 to 857~GHz and scanned 
the CMB sky at much higher angular resolution, from 10 to $5\arcmin$. 
Due to its large frequency coverage, high resolution and low noise, 
the {\it Planck} satellite is an optimal instrument for detecting the TSZ 
distortion of the CMB spectrum induced by clusters.
To clean the CMB data from foreground contributions, the Planck Collaboration also 
released maps of (a) thermal dust and residual Cosmic Infrared Background emission,
(b) synchrotron, free-free and spinning dust emission, (c) CO contributions,
that are most important at 100, 217 and 353~GHz and (d) maps of dust temperature
and opacity \citep{planck_12}. Additional products 
required to analyze the data such as masks and noise maps were also released.
Using {\it Planck} data and a sample of 813 clusters up to redshift $z\sim 1$,
\cite{hurier} obtained $\alpha=0.009\pm 0.017$. Their sample
included X-ray and TSZ selected clusters. Using clusters detected with
the South Pole Telescope, \cite{spt} measured $\alpha=0.017^{+0.030}_{-0.028}$, 
also compatible with adiabatic evolution. 

In this article we will apply the techniques developed in \cite{demartino}
to test the standard temperature-redshift relation. Our cluster catalog 
comprises $782$ clusters selected from the X-ray ROSAT data and with 
well measured properties; of those, 481 will be used in this study. The outline
of this paper is as follows: In Sec.~2 we describe our data and pipeline; in Sec.~3 
we present our results and in Sec.~4 we summarize our main conclusions. 

\section{Data and data processing}

We constrain the temperature-redshift evolution of the background temperature
using a cluster catalog selected from ROSAT data and the
{\it Planck} Nominal maps released in  2013 \footnote{{\it Planck}
data can be downloaded from \url{http://www.cosmos.esa.int/web/planck}}.
We will only use HFI data, originally released in  
Healpix format with resolution $N_{side}=2048$ \citep{healpix}. 

\subsection{Cluster Sample.}\label{sec:xray}

\begin{figure}
\centering 
\epsfxsize=.9\textwidth \epsfbox{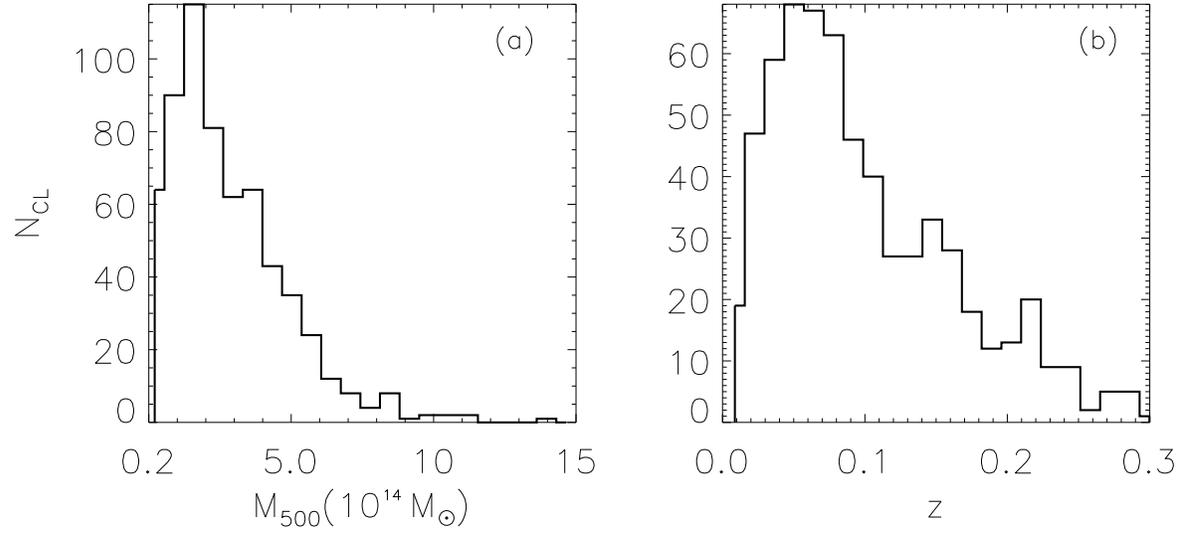}
\vspace*{-4.cm}
\caption{Histograms with the mass (a) and redshift (b) distributions of 
the 481 clusters used in this study.
}
\label{fig1}
\end{figure}

Our catalog contains 782 clusters, of which 623 are outside the Planck 
PCCS-SZ-Union mask \citep{planck_28,planck_29} (see below). 
They have been selected from three X-ray flux limited surveys: the 
ROSAT-ESO Flux Limited X-ray catalog (REFLEX, B\"ohringer et al. 2004), 
the extended Brightest Cluster Sample (eBCS, Ebeling et al. 1998 and 2002)
and the Clusters in the Zone of Avoidance (CIZA, Ebeling et al. 2002).
Details of the combined catalog are given in \citet{kocevski06}.
All clusters have well measured positions, X-ray fluxes and luminosities 
in the [0.1,2.4]KeV ROSAT band, spectroscopic redshifts
and angular extents of the X-ray emitting region. 
The catalog also lists the X-ray temperature derived from the $L_X-T_X$ relation 
of \citet{white} and the core radii $r_c$ and central electron density $n_{e,0}$
obtained by fitting a $\beta=2/3$ model to the ROSAT data. 
We compute the radius at which the mean overdensity of
the cluster is 500 times the critical density, $r_{500}$, from the 
$r_{500}-L_X$ relation of \citet{bohringer2}. We define $M_{500}$ as 
the mass enclosed within a sphere of radius $r_{500}$ and 
the angular size $\theta_{500}=r_{500}/d_A$, where
$d_A(z)$ is the diameter angular distance of each cluster
in the $\Lambda$CDM model. In Fig. \ref{fig1} we represent
the redshift and mass distribution of the clusters in our catalog.

The clusters in our sample have masses in the interval 
$M_{500}=[0.2-10]\times 10^{14}$M$_\odot$ and are located at 
redshifts $z\le 0.3$, relatively lower compared with other catalogs. 
The catalog used by \cite{hurier} contains X-ray and Planck
selected clusters. Compared with our catalog,
the MCXC sample \citep{piffaretti} includes clusters from sources that 
we have not considered such as NORAS \citep{noras}, SGP \citep{sgp} and NEP 
\citep{nep}. Also, our catalog does not include clusters from X-ray pointed surveys 
which tend to find mainly low-mass systems but out to high redshift.
For redshifts $z<0.3$ and masses $M_{500}\ge 10^{14}M_\odot$
all our clusters except ten are listed in the MCXC sample.
In the same mass and redshift range the Planck SZ catalog contains 
555 clusters. Of those, only 239 are at less than $10'$ away from 
members of our sample.
An important difference with the latter catalog is that
the Planck Collaboration estimates $M_{500}$ using the scaling
relation $M_{500}-Y_{X,500}$ 
of \cite{arnaud} while we use the $r_{500}-L_X$ relation from
\cite{bohringer2} and the resulting masses differ, on average, by 30\%.
\cite{spt} used a sample of 158 clusters with redshifts $z\le 1.35$, 
selected by their TSZ signal. When cluster candidates are
identified using CMB data the adiabatic evolution is assumed, biasing
cluster selection towards those candidates that mimic this behavior. 
The South Pole group verified that this effect was not significant.
Nevertheless, due to their selection criteria, negative temperature fluctuations
were more likely negative than positive and taking into account this second effect
increased their error bars a 30\% \citep{spt}.
Therefore, it is important to verify the results of other groups using
only X-ray selected clusters since they will not be affected by these biases.
In Table~\ref{table1} we present the mean redshift, angular scale,
X-ray luminosity and mass
averaged over the full cluster sample and in several subsamples selected 
according to X-ray luminosity mass and redshift.  For better comparison
we chose the redshift bins as in \cite{hurier}; within each redshift bin, 
we select all clusters with $M_{500}\ge 2\times 10^{14}M_\odot$ or
$L_X\ge 2.5\times 10^{44}$ erg/s. In the table we do not quote a redshift bin in
the mass and X-ray luminosity subsamples when it coincides with the 
bin of the full sample. 

\begin{table}[!ht]
\begin{center}
\begin{tabular}{|c|c|c|c|c|c|}
\hline
& & & & & \\[0.8mm]
Subset & Ncl & $\bar{z}$  & $\bar{\theta}_{500}$ & $\bar{L}_X$ & $\bar{M}_{500}$ \\
       &     &            &   (arcmin)      & ($10^{44}$erg/s) & ($10^{14}M_\odot$)\\
\hline
\textbf{All Clusters}  & 481 & 0.106  &  12.3  &  2.35  & 3.1  \\[0.8mm]
\hline
$0.0<z<0.05$  & 32  & 0.035  &  21.2  &  0.51  &  1.2 \\[0.8mm]
$0.05<z<0.10$ & 186 & 0.074  &  12.5  &  1.34  &  2.4 \\[0.8mm]
$0.10<z<0.15$ & 114 & 0.123  &  8.82  &  2.20  &  3.3 \\[0.8mm]
$0.15<z<0.20$ & 83  & 0.169  &  7.64  &  4.17  &  4.8 \\[0.8mm]
$0.20<z<0.25$ & 46  & 0.222  &  6.58  &  6.13  &  5.8 \\[0.8mm]
$0.25<z<0.30$ & 20  & 0.274  &  6.15  &  10.1  &  7.6 \\[0.8mm]
\hline
${\bf{M_{500} \ge 2\times10^{14}M_\odot}}$&397 &0.134 &10.49 &3.40 &4.1 \\[0.8mm]
\hline
$0.0<z<0.05$   & 20  & 0.039  &  26.5  &     1.67  &    3.1 \\[0.8mm]
$0.05<z<0.10$  & 121 & 0.078  &  13.4  &     2.01  &    3.2 \\[0.8mm]
$0.10<z<0.15$  & 107 & 0.124  &  8.85  &     2.28  &    3.4 \\[0.8mm]
\hline
${\bf{L_X\ge 2.5\times10^{44}}}$erg/s & 201 & 0.152 & 13.51 & 5.32 & 5.7\\[0.8mm]
\hline
$0.0<z<0.05$  & 3  & 0.037  &  38.4  &    3.47&    5.2 \\[0.8mm]
$0.05<z<0.10$ & 25 & 0.077  &  16.6  &    4.28&    5.6 \\[0.8mm]
$0.10<z<0.15$ & 36 & 0.129  &  9.59  &    3.51&    4.6 \\[0.8mm]
$0.15<z<0.20$ & 71 & 0.171  &  7.72  &    4.51&    5.1 \\[0.8mm]
\hline
\end{tabular}
\caption{Average properties of different cluster subsamples from the 
general catalog. Clusters have been selected by luminosity and mass within 
each redshift bin.}
\label{table1}
\end{center}
\end{table}

\subsection{Foreground Cleaned Planck Nominal Maps. \label{sec:2.2}}

The nine frequency maps released by the Planck Collaboration in 2013
contained foreground emissions in addition to the intrinsic CMB temperature 
anisotropies and the instrumental noise. Since the different frequencies
have different angular resolutions, we bring all maps to the common resolution
of $10\arcmin$, the lowest of the HFI channels. Due to their lower resolution 
and higher instrumental noise we will not use the LFI data. We remove the cosmological 
and KSZ anisotropies by subtracting the LGMCA CMB map from \citet{LGMCA1,LGMCA2} 
smoothed to the same $10\arcmin$ resolution. This map was constructed from 
the latest {\it Wilkinson 
Microwave Anisotropy Probe} 9yr and first {\it Planck} 2013 data releases.
The data were combined using a component separation method based on 
the sparsity of the foregrounds in the wavelet domain that requires frequency
information to remove the TSZ effect. The resulting map
has low dust contamination at $\ell<1000$ and the KSZ anisotropy is preserved 
while the TSZ signal is removed. In this respect, it is more useful than the 
{\it Planck} reconstructions of the CMB anisotropies, like SMICA, that contain 
substantial TSZ residuals. Next, we clean the foreground emission around each 
cluster. At frequencies higher than 100~GHz, thermal dust emission dominates 
over most of the sky \citep{planck_12}. This contribution is commonly described 
as a modified black-body spectrum with a power-law emissivity 
$\epsilon_\nu\propto \nu^{\beta_d}$ with a
slope $\beta_d\approx 1.5-1.8$ \citep{planck_11}, independent of frequency
but varying across the sky even on scales as small as $5\arcmin$. The
dust contribution is largest at 857~GHz and dominates over the 
cosmological signal. Then, we can use this channel as a template for thermal dust 
to clean patches $P(\nu,{\bf x})$ centered on each cluster position ${\bf x}$ 
and on each maps of frequency $\nu$. When the 
dust emission parameters and frequency dependence are known, one
can generate dust templates at different frequencies to be subtracted
from the data. In this paper we will use a different approach,
proposed by \citet{diego_02}, that minimizes the contribution to the map
at frequency $\nu$ of all the foregrounds that correlate with the 857~GHz channel.
Specifically, the weight $w(\nu)$ minimizes the
difference $[P(\nu,{\bf x})-w(\nu)P(857~{\rm GHz},{\bf x})]^2$.
Weights are computed on a ring around the cluster region,
defined as $\mathcal{R}=[\theta_{cl},\theta_{patch}]$, and are given by
\begin{equation}
 w(\nu)=\frac{\Sigma_{\bf x\in \mathcal{R}}P(\nu,{\bf x})P({\rm 857~GHz},{\bf x})}
{\Sigma_{\bf x\in \mathcal{R}}[P({\rm 857~GHz},{\bf x})]^2} .
\label{eq:weight}
\end{equation}
To avoid overlapping with the TSZ emission from the clusters in our
catalog, we masked
a region of $3\theta_{500}$ around them. Excising 
these pixels changes the weight by less than 2\%. We assume that the dust temperature 
and spectral index are constant within the ring. This approximation is
more accurate for the less extended clusters. Ring sizes vary 
depending on the cluster extent: $\theta_{cl}=2\theta_{500},\; 
\theta_{patch}=3\theta_{500}$ for all clusters with $\theta_{500}\ge 20\arcmin$,
otherwise $\theta_{cl}=3\theta_{500}$ and $\theta_{patch}=[5-7]\theta_{500}$.
If we would use a ring of the same size for all clusters, the patch would be
too large for the most extended clusters and the approximation
of constant temperature and constant spectral index will not
held. On the contrary, if $\theta_{500}$ is too small, then the patch would
contain few pixels and the statistical weight would be unreliable. We checked
that the above angular extents yield the smallest foreground residuals around
clusters. The procedure is robust since weights did not change 
if $\theta_{patch}$ and $\theta_{cl}$ varied by a factor of two. Then,
even if the smallest/largest annuli could contain a small TSZ contribution 
from the cluster or nearby clusters, the weights were not affected. 

The first three CO rotational transition lines 
$J=1\rightarrow 0$, $J=2\rightarrow 1$, and $J=3\rightarrow 2$ 
at 115, 231 and 346~GHz, respectively, present the largest transmission 
coefficients making them a significant foreground component in the 
Planck intensity maps. We corrected this contribution using the estimated
CO emission maps provided by the Planck Collaboration. Three type of CO maps
has been made available \citep{planck_13}: Type 1 maps are available for the three
frequencies but are too noisy and are sensitive only to the brightest
regions on the galactic plane to be useful.
Type 2 maps are less noisy but they are only available 
at 100 and 217~GHz. We corrected these two frequencies and found the
correction to affect very little the final cleaned patches. Since the 
transition at 353~GHz is the weakest of the three, we expect that the CO 
contamination at 353~GHz will neither produce a significant effect 
on the final results.

To reduce the contamination from point sources and foreground residuals near 
the galactic plane, we used the PCCS-SZ-Union mask \citep{planck_28,planck_29}. 
This mask was constructed using  the {\it Planck} Catalogue of Compact Sources 
(PCCS). It is the union of six masks, 
one for each HFI channel, and the mask of the Galactic Plane and the Magellanic Clouds.
Our cleaning method does not give satisfactory results for the faintest clusters, 
so we restricted our analysis to those clusters with a X-ray luminosity in 
the ROSAT [0.1-2.4]KeV band of $L_X \ge 0.5\times10^{44}$erg/s, reducing the 
total number of clusters to 481.

\begin{figure}
\centering 
\epsfxsize=.8\textwidth \epsfbox{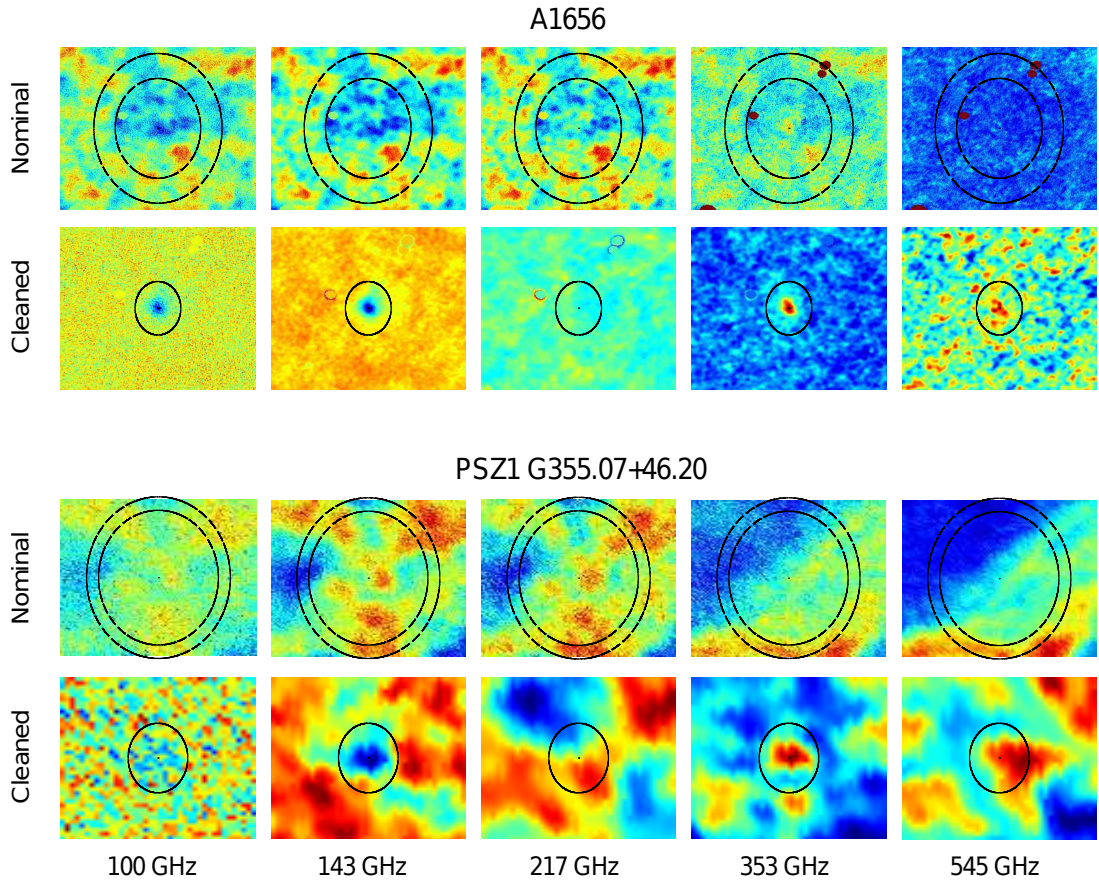}
\caption{Original Planck Nominal and foreground cleaned patches
centered on the position of the Coma [A1656] and the PSZ1 G355.07+46.20 clusters at
the HFI frequencies 100-545~GHz. 
Blue/red colors correspond to negative/positive temperature fluctuations.
For Coma, the patch is $7^0\times 7^0$ in size
and the temperature range is $[-300,300]\mu$K; for PSZ1 G355.07+46.20 
the size of the patch is $2.5^0\times 2.5^0$ in the
Nominal and $1^0\times 1^0$ in the Cleaned maps and the
temperature is within $[-200,200]\mu$K. In the Nominal maps, the inner and outer
circles have radius $\theta_{cl}$ and $\theta_{patch}$, respectively; in
the Cleaned maps, the angular radius is $\theta_{500}$ for the given cluster.}
\label{fig2}
\end{figure}

To illustrate how effectively our pipeline removes foregrounds and the differences
in the final results between low and extended and high and compact 
redshift clusters, in Fig.~\ref{fig2} we show the 
temperature fluctuation on patches of angular size $75^\prime\times 75^\prime$ for 
two clusters: Coma, with an angular extent of $\theta_{500}=48.1'$ and redshift $z=0.023$,
and PSZ1 G355.07+46.20 with $\theta_{500}=9.2'$ and $z=0.21$. The first
and third rows correspond to the temperature anisotropies on the original
{\it Planck} Nominal maps at frequencies $100-545$~GHz. 
The second and fourth rows show the same regions on the foreground 
cleaned maps.  In Coma and PSZ1 G355.07+46.20
we fixed the temperature range to be $[-300,300]\mu$K and $[-200,200]\mu$K,
respectively. In the Planck nominal maps, the two circles have a angular radius of
$\theta_{cl}$ (inner circle) and $\theta_{patch}$ (outer circle). In the
foreground clean maps the radius corresponds to $\theta_{500}$.
The TSZ temperature anisotropy is negative at 100 and 143GHz;
at 217~GHz, it is greatly reduced and it changes sign at higher frequencies.
The data at 545~GHz is dominated by foreground residuals and our pipeline produces 
reliable estimates of the TSZ signal only for the brightest clusters. At 
this frequency, foreground residuals dominate over cluster anisotropies and 
we could not use it to determine/constrain deviations from adiabatic evolution.

\subsection{Testing our Cleaning Procedure}\label{sec:2.3}

\begin{figure}
\centering
\epsfxsize=.99\textwidth \epsfbox{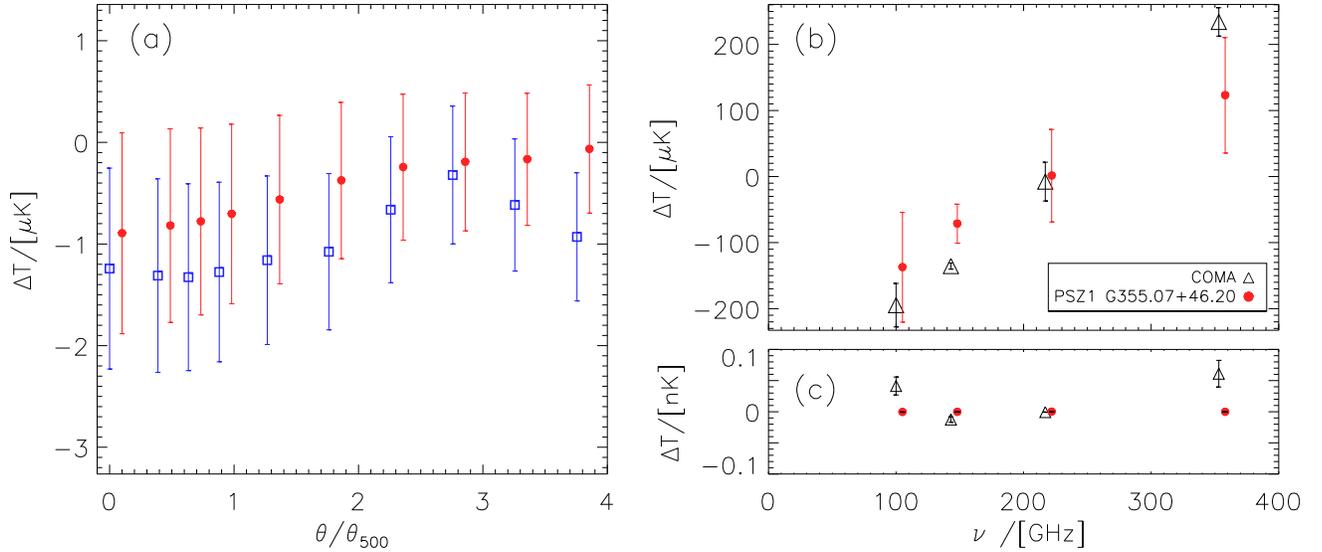}
\vspace*{-4.5cm}
\caption{(a) Mean temperature anisotropy and its error
evaluated on the LGMCA map on discs of different radius, in units of the
$\theta_{500}$.  Blue open squares, red solid circles 
correspond to the average at the real position of the 481 real clusters 
and at the same number of random locations on the sky, respectively. 
The error bars represent the error on the mean. (b) TSZ anisotropy, 
measured on a disc of size $\theta_{500}$ on our CMB removed, foreground cleaned data
as a function of frequency for the two clusters represented in Fig. \ref{fig2}. 
Red circles have been shifted for to facilitate their view. (c) Average temperature 
anisotropy evaluated in the ring $[\theta_{cl}, \theta_{patch}]$ for the
same two clusters. 
The error for PSZ1 G355.07+46.20 is too small to be seen in the figure.}
\label{fig3}
\end{figure}

We checked that the LGMCA map does not contain significant
TSZ residuals by computing the mean temperature anisotropy at the 
position of our clusters on discs of different sizes, in units of $\theta_{500}$.
In Fig.~\ref{fig3}a, blue squares and red solid circles represent
the mean at the actual position of our 481 clusters and 
the mean at 481 random position in the sky, respectively.
The error bars are the error on the mean estimated from 100 simulations.
For a better view, the
results are displayed slightly shifted. The averages at the cluster positions
are marginally biased towards negative values compared with a random 
distribution. For apertures $\ge\theta_{500}$, this bias is $\sim -1\mu$K 
or smaller and well within the error bar. Since the TSZ is removed from
the LGMCA map using the standard ($\alpha=0$) frequency dependence of
the effect, this offset could be systematic and 
not the mean of some random residuals left at the cluster locations
by the component separation process. If $\alpha \ne 0$, the TSZ residuals that 
would remain would change with frequency. If the offset were systematic
the TSZ effect would be shifted upwards by $1\mu$K after subtracting the LGMCA 
map, modifying the dependence with $\alpha$ but not removing it.
We will discuss this point further in Sec.~3.3.

To illustrate that the TSZ anisotropy has been preserved in our foreground 
cleaned and CMB subtracted maps, in Fig.~\ref{fig3}b we plot
the mean temperature anisotropy averaged on a disc of radius $\theta_{500}$. 
Open triangles correspond to the profile of the Coma cluster, while solid 
circles to PSZ1 G355.07+46.20. The error bars were computed by placing discs
with the same angular extent at one thousand random positions outside the
clusters in our sample; they are much smaller than those
of Fig~\ref{fig3}a since only the instrumental noise
and foreground residuals contribute. The difference
in the errors between both clusters are due to their different angular extent. 
Third, we verified that there are no TSZ anisotropy on the rings selected to
compute the weights of eq.~(\ref{eq:weight}) that could potentially bias the result.
Fig.~\ref{fig3}c shows that the mean temperature anisotropy on the rings around
of the two clusters is $\sim 10^{-4}\mu$K, 
negligible compared with the errors on the cluster profiles;
around other clusters the mean was always $\le 10^{-3}\mu$K, 
also negligible. To conclude, subtracting the 
LGMCA map from {\it Planck} Nominal maps effectively removes the cosmological
CMB and KSZ signals while preserving the TSZ anisotropy and its dependence
with $\alpha$.

\subsection{Error Bar Estimation}

For each cluster configuration, error bars were computed by evaluating the mean 
temperature fluctuation on 1,000 random positions in foreground cleaned maps
on a disc with the same angular extent than the clusters in any given subsample. 
The random positions were chosen to be at least 2 degrees away from the 
location of our clusters to guarantee that the random discs will never
overlap with them. We removed the foreground contamination using 
the same procedure than at cluster locations. We verified that 
the mean of the simulations was compatible with zero. 
The correlation matrix between different frequencies was
computed by averaging over all simulations 
\begin{equation}
C(\nu_i,\nu_j)=\frac{\langle [\delta T(\nu_i)-\mu(\nu_i)]
[\delta T(\nu_j)-\mu(\nu_j)]\rangle}{\sigma(\nu_i)\sigma(\nu_j)} ,
\label{eq:correlation}
\end{equation}
where $\mu(\nu_i)=\langle\delta T(\nu_i)\rangle$, and 
$\sigma(\nu_i)=\langle [\delta T(\nu_i) -\mu(\nu_i)]^2\rangle^{1/2}$.
To compare with previous work, we repeated the process
and computed the correlation matrix between foreground cleaned maps before
subtracting the LGMCA map using our full catalog and compared it with the correlation 
matrix of \citet{hurier}, obtained with a different cluster catalog. 
While the latter authors used a fix aperture of $20\arcmin$ for all clusters
we chose apertures of size $\theta_{500}$ for each cluster.  We 
found that our off-diagonal terms were slightly smaller; the 
difference between the two correlation matrices differed between 2\% for 
the element C(100GHz,143GHz) to less than 10\% for C(100GHz,353GHz), 
allowing us to conclude that our cleaning 
procedure was comparatively as effective as theirs.

\section{Results}

To estimate/constrain the temperature-redshift relation we used the ratio 
and fit methods described in the introduction. We derived an X-ray
temperature for all clusters in our catalog from the measured X-ray
luminosity using the $L_X-T_X$ relation of \cite{white}. 
The cluster temperature allow us to include relativistic corrections
\citep{itoh1998,nozawa2006}.  If we denote $\theta_e=(k_BT_e/m_ec^2)$, 
the frequency dependence of the TSZ effect
including relativistic corrections up to fourth order
in the cluster temperature, $G_{4}(\nu)$, can be written as
\begin{equation}
G_{4}(\nu)=G(\nu)\left[1+\theta_e\left(\frac{Y_1}{Y_0}\right)
+\theta_e^2\left(\frac{Y_2}{Y_0}\right)^2
+\theta_e^3\left(\frac{Y_3}{Y_0}\right)^3
+\theta_e^4\left(\frac{Y_4}{Y_0}\right)^4\right]
\end{equation}
where $Y_0,...,Y_4$ are defined in eqs~2.24-2.29 
of \cite{nozawa1998}. The correction is relatively small;
for the most massive clusters it amounts to a few percent. 
Corrections depending on the {\it unknown} cluster peculiar velocities
were not included but for the values expected in the concordance
model, they are also negligible. Hereafter, in order
to simplify the notation we will drop the subindex; it should be understood that
relativistic corrections were included except when indicated otherwise.
Finally, let us mention that as
a prior we considered $\alpha=[-1,1]$, subdivided in 2001 equally spaced steps. 

\subsection{Ratio method}

As indicated in Sec.~\ref{sec:2.2}, all maps were reduced to a common
resolution of $10\arcmin$ to eliminate the differences in the angular resolution.
In this statistic, the ratio of the mean temperature anisotropy
on a disc of a fixed angular size $\theta_{cl}$ at two frequencies,
$\overline{\delta T}(\nu_1)/\overline{\delta T}(\nu_2)$, does not depend on
the cluster pressure profile. In the absence of noise, this ratio would be
equal to the theoretical value at each redshift
$R(\nu_1,\nu_2,\alpha)=G(\nu_1,\alpha)/G(\nu_2,\alpha)$,
providing a direct measurement of $\alpha$. However, instrumental
noise and foreground residuals complicate the analysis.
As discussed in \citet{luzzi}, for each cluster $j$ 
the probability distribution of the ratios, $\mathcal{P}_j(R)$, is 
\begin{equation}
\mathcal{P}_j(R(\nu_1,\nu_2,\alpha))=\frac{1}{2\pi\sigma_{\nu_1}\sigma_{\nu_2}}
\times \int_{-\infty}^\infty x \exp \Biggl(-\Biggl[
\frac{[x-\overline{\delta T}(\nu_1)]^2}{2\sigma_{\nu_1}^2}+
\frac{[xR(\nu_1,\nu_2,\alpha)-\overline{\delta T}(\nu_2)]^2}{2\sigma_{\nu_2}^2}
\Biggr]\Biggr)dx ,
\label{eq:ratio_probability}
\end{equation}
where $\sigma_{\nu}$ is the error on mean temperature at the
cluster location $j$ at frequency $\nu$. The likelihood function
is ${\cal L}\propto \Pi_{j}P_j$; we did not include the correlation 
between the different frequencies so our analysis will underestimate the errors.
We took five ratios between 
frequencies from 100~GHz to 353~GHz: 100/353, 143/353, 217/353,
100/143, 217/143. We chose in the denominator the
channel with the smallest instrumental noise and excluded 217~GHz
to avoid dividing by zero when $\alpha\simeq 0$.
Temperature anisotropies were evaluated on discs of radii $\theta_{cl}=\theta_{500}$.
Following \citet{hurier}, we divided our sample in six redshift bins
of equal width $\Delta z=0.05$; then, in eq.~(\ref{eq:ratio_probability})
$\overline{\delta T}(\nu_1)$ and  $R(\nu_1,\nu_2,\alpha)$ are now the averages 
over all the clusters in the bin. 

\begin{figure} 
\epsfxsize=.9\textwidth \epsfbox{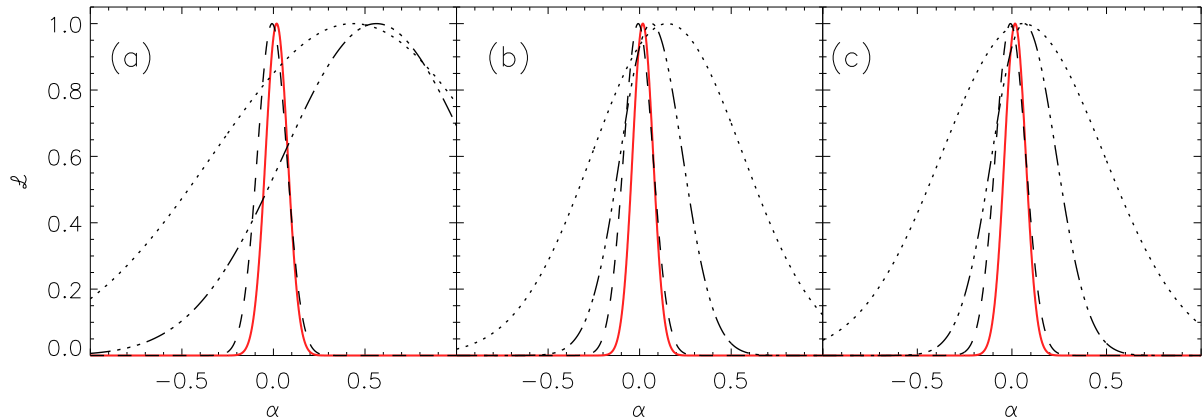}
\vspace*{-4.5cm}
\caption{Ratio method: (a) likelihoods corresponding to 201 
clusters with $L_X \ge 2.5\times10^{44}$erg/s, in (b)
to 397 clusters with $M_{500} \ge 0.2\times10^{15}M_\odot$ and 
in (c) to the full sample of 481 clusters. 
Dotted, dash-dotted and dashed lines correspond to  
clusters in the interval $z=[0.05, 0.1]$, 
$z=[0.15, 0.2]$ and  $[0.25, 0.3]$, respectively. 
The red solid lines correspond to the full likelihood. Anisotropies
were taken as averages on discs of size $\theta_{500}$.}
\label{fig4}
\end{figure}

In Fig.~\ref{fig4} we plot the likelihood function for different cluster
subsets: In (a), we considered the 201 most X-ray luminous clusters, with 
X-ray luminosity $L_X \ge 2.5\times10^{44}$erg/s, in (b) the 397 clusters with 
mass $M_{500} \ge 2\times10^{14}M_\odot$ and, finally, in (c) we 
plot the likelihood of our full cluster sample. For illustration, only the
likelihood of three redshift bins are plotted. In
each panel, the dotted, dash-dotted and dashed lines correspond to the
bins $[0.05, 0.1]$, $[0.15, 0.2]$, $[0.25, 0.3]$, respectively, and
the solid lines correspond to the full sample. The measured values of $\alpha$ 
are given in Table~\ref{table2}; within each redshift bin we computed
$\alpha$ in different subsamples, with clusters selected in
X-ray luminosity and mass as indicated in Table~\ref{table1} in order
to test the relative contribution of the different cluster subsamples to the 
final error budget.
In all redshift bins, the results are always compatible with zero. 
Notice that our error bars do not necessarily scale with the square root of
number of clusters since, as indicated in Sec.~2.2, 
our cleaning procedure, and consequently the amplitude of
the foreground residuals, depend on the 
cluster extent. We checked that the results were 
very similar if the temperature averages were taken on discs of radii 
$\theta_{cl}=2\theta_{500}$. Using the full sample, we constrain
the deviation from  adiabatic invariance to be $\alpha=-0.03\pm 0.06$. 
This estimate is a factor of 2 worse than the result obtained 
by the SPT group \citep{spt},  even with our underestimated
error bars. This is to be expected since
their maps have a resolution of $1.5\arcmin$ and even if their sample is
smaller in size their clusters are, on average, at higher redshifts where
the effect of the non-adiabatic evolution is most noticeable.

\begin{table}[!ht]
\begin{center}
\begin{tabular}{|c|ccc|ccc|ccc|}
\hline
Subset & $N_{cl}$ & $\alpha_{L_X}$ & $\sigma_{\alpha_{L_X}}$  
& $N_{cl}$ &$\alpha_{M_{500}}$ & $\sigma_{\alpha_{M_{500}}}$ 
& $N_{cl}$ & $\alpha_{\theta_{500}}$ & $\sigma_{\alpha_{\theta_{500}}}$ \\
\hline
     All      & 201 & 0.02  & 0.06 & 397 & 0.02 & 0.06 & 481 & 0.02  & 0.06\\
$0.0<z<0.05$  &   3 & 0.13  & 1.0  &  20 &-0.05 & 0.74 &  32 & 0.03  & 0.67\\
$0.05<z<0.10$ &  25 & 0.44  & 0.77 & 121 & 0.15 & 0.41 & 186 & 0.06  & 0.43\\
$0.10<z<0.15$ &  36 & 0.95  & 1.27 & 107 & 0.32 & 0.34 & 114 & 0.26  & 0.29\\
$0.15<z<0.20$ &  71 & 0.56  & 0.51 &  83 & 0.07 & 0.17 &  83 & 0.07  & 0.17\\
$0.20<z<0.25$ &  46 & 0.01  & 0.01 &  46 & 0.01 & 0.01 &  46 & 0.01  & 0.01\\
$0.25<z<0.30$ &  20 &-0.01  & 0.08 &  20 &-0.01 & 0.08 &  20 &-0.01  & 0.08\\
\hline
\end{tabular}
\caption{Values of $\alpha$ estimated using the ratio method.
Temperatures are averaged on discs of radius $\theta_{500}$ over all the
clusters in the redshift bin.} \label{table2}
\end{center}
\end{table}

\subsection{Frequency fit method}\label{sec:fitmethod}

As an alternative method, we constrain the adiabatic evolution of 
the Universe by fitting the frequency dependence of the TSZ anisotropy,
$\Delta  T(\hat{n})=T_0\bar{Y}_cG(\nu,\alpha)$. 
The Comptonization parameter $\bar{Y}_C$ is the average on a disc
of radius $\theta_{500}$. We follow \cite{hurier} and take it as a free parameter.
In this case, both $\bar{Y}_C$, and $\alpha$ are fit to the data. 
We took the flat prior $\bar{Y}_C=[0,300]\mu$K divided
in intervals of $\Delta \bar{Y}_C=0.15\mu$K.  Since all maps have the same
resolution, $\bar{Y}_C$ is independent of frequency. We compute the 
likelihood function as
\begin{equation}
-2\log{\cal L}=\sum_{i,j} \sum_{k=1}^{N_{cl}}
\left[{\overline{\delta T_k}(\nu_i)-\bar{Y}_{C,k}G(\nu_i,\alpha)}\right]
C^{-1}(\nu_i,\nu_j)\left[\overline{\delta T_k}(\nu_j)-\bar{Y}_{C,k}G(\nu_j,\alpha)\right] ,
\label{eq:chisq_fit}
\end{equation}
where $C(\nu_i,\nu_j)$, given in eq.~(\ref{eq:correlation}), is
\begin{equation}
C(\nu_i,\nu_j)=\begin{pmatrix}
1.0000  &  0.9258  &  0.4603  &  0.1435 \\
0.9258  &  1.0000  &  0.4995  &  0.3168 \\
0.4603  &  0.4995  &  1.0000  &  0.3605 \\
0.1435  &  0.3168  &  0.3605  &  1.0000 \\
\end{pmatrix}, 
\end{equation}
where $\nu_i=(100,143,217,353)$GHz. Notice that by using the LGMCA map 
to subtract the intrinsic CMB signal we have significantly reduced the
correlation between frequencies and the variance of the foreground
clean maps compared to \cite{hurier}.

\begin{figure}
\epsfxsize=.9\textwidth \epsfbox{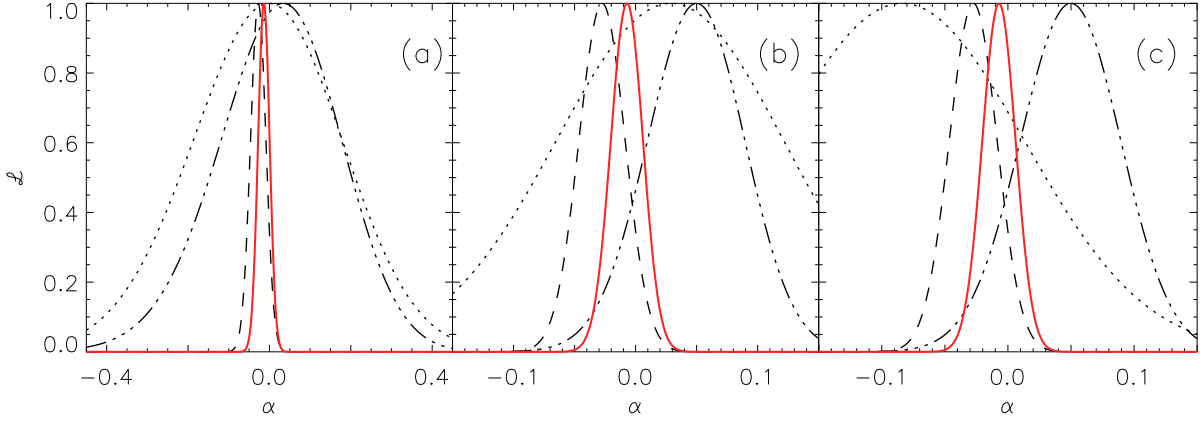}
\vspace*{-4.5cm}
\caption{Fit method: Panels represent the same
likelihoods than in Fig.~\ref{fig4} and lines follow the same convention.
}
\label{fig5}
\end{figure}

\begin{figure}
\epsfxsize=.9\textwidth \epsfbox{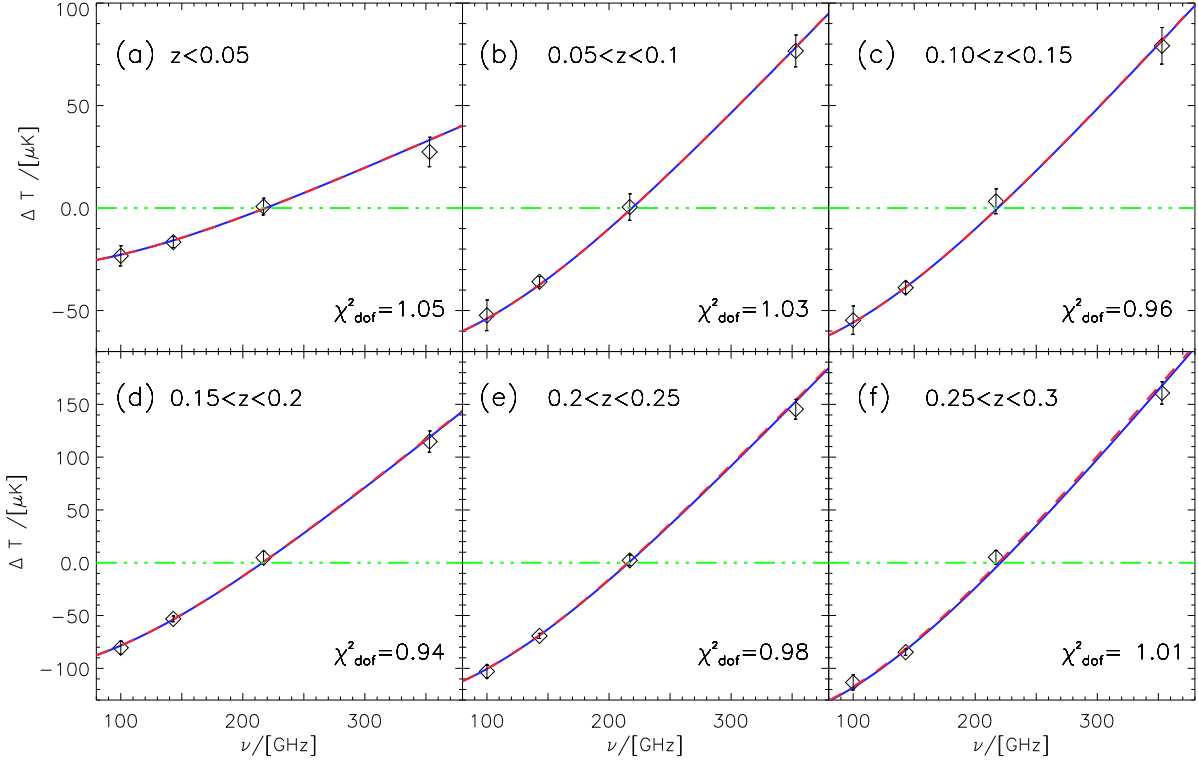}
\caption{Frequency dependence of the TSZ temperature anisotropies
averaged over the cluster in the redshift subsamples. Solid lines correspond
to the standard adiabatic evolution, $\alpha=0$, without relativistic
corrections while dashed lines include these corrections.}
\label{fig6}
\end{figure}

In Fig.~\ref{fig5} we present the likelihood function, 
marginalized over the Comptonization parameter, for the same cluster
subsamples and redshift bins than in Fig.~\ref{fig4}, with lines following
the same convention. The numerical results are given in Table~\ref{table3};
in this case we find that $\alpha=-0.007 \pm 0.013$, compatible with zero.
We verified that changing the prior to $\bar{Y}_C=[-300,300]\mu$K,
that would allow for foreground residuals to change the sign of the TSZ effect, 
did not modify the error bars, indicating that the data prefers negative
values in the Rayleigh-Jeans and positive values in the Wien part of the
spectrum. To illustrate the accuracy of our results,
in Fig.~\ref{fig6} we plot the average temperature anisotropies of
all the clusters the redshift bins of Table~\ref{table1} and their errors,
for the different frequencies.
The solid line corresponds to the best fit $\bar{Y}_CG(\nu,\alpha)$ without
relativistic corrections and 
the dashed line to the best fit including these corrections.
Notice that only at $z>0.2$ this correction can be distinguished
from the non-relativistic effect. We also indicate the $\chi^2$
per degree of freedom of the best fit model. 

\begin{figure}
\epsfxsize=.9\textwidth \epsfbox{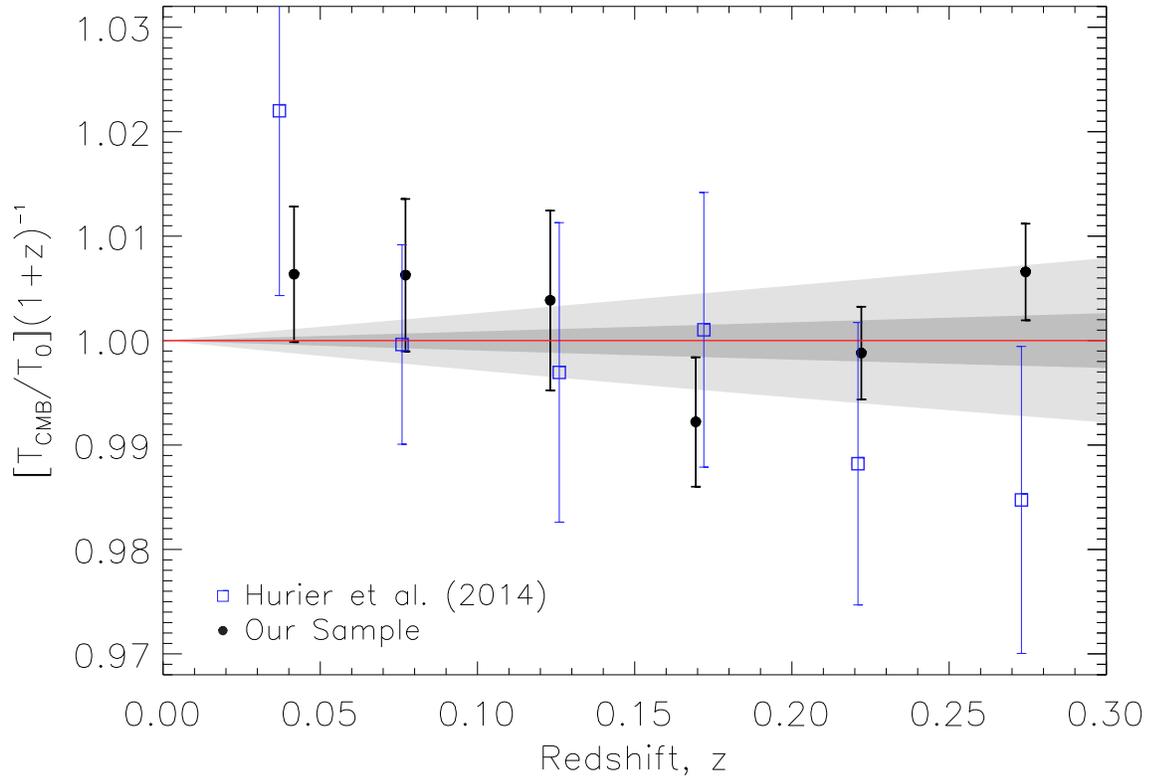}
\caption{Constraints on the adiabatic evolution of the background
temperature at different redshifts  given in Fig.~\ref{fig6}. 
The results of Table~\ref{table3} are represented by solid circles 
with their associated error bars. For comparison, open 
squares and error bars correspond to the results of \citet{hurier}. 
Points are displayed at the mean redshift of the clusters in the bin.
The shaded areas correspond to the 68\% and 99.7\% confidence level.
}
\label{fig7}
\end{figure}

Our result represents a 30\% improvement over \citet{hurier} and it was obtained
with a smaller sample of clusters located at a lower mean redshift. 
In Fig.~\ref{fig7} we present a more thorough comparison. The shaded regions
correspond to the $1$ and $3\sigma$ errors given in Table~\ref{table3}.
Notice that in the same redshift range, the measured values
are in an excellent agreement and are fully compatible with an adiabatic
evolution of the Universe, the main difference being that our error 
bars are significantly smaller. The reduction on the uncertainty is due 
to using an accurate CMB template to subtract the cosmological signal.

\begin{table}[!ht]
\begin{center}
\begin{tabular}{|c|ccc|ccc|ccc|}
\hline
Subset & $N_{cl}$ & $\alpha_{L_X}$ & $\sigma_{\alpha_{L_X}}$  
& $N_{cl}$ &$\alpha_{M_{500}}$ & $\sigma_{\alpha_{M_{500}}}$ 
& $N_{cl}$ & $\alpha_{\theta_{500}}$ & $\sigma_{\alpha_{\theta_{500}}}$ \\
\hline
All           & 201 &-0.013  & 0.014 & 397 &-0.006 & 0.013 & 481 &-0.007 & 0.013\\
$0.0<z<0.05$  &   3 &-0.27   & 0.23  &  20 &-0.21  & 0.17  &  32 &-0.16  & 0.16\\
$0.05<z<0.10$ &  25 & 0.00   & 0.18  & 121 & 0.03  & 0.09  & 186 &-0.08  & 0.09 \\
$0.10<z<0.15$ &  36 & 0.08   & 0.20  & 107 &-0.01  & 0.09  & 114 &-0.03  & 0.07 \\
$0.15<z<0.20$ &  71 & 0.03   & 0.05  &  83 & 0.05  & 0.04  &  83 & 0.05  & 0.04\\
$0.20<z<0.25$ &  46 & 0.01   & 0.02  &  46 & 0.01  & 0.02  &  46 & 0.01  & 0.02\\
$0.25<z<0.30$ &  20 &-0.03   & 0.02  &  20 &-0.03  & 0.02  &  20 &-0.03  & 0.02\\
\hline
\end{tabular}
\caption{Estimated values of $\alpha$ and their uncertainties using the fit method.
} \label{table3}
\end{center}
\end{table}

We can use the upper limit on $\alpha$ given above to constrain
the phenomenological model of \cite{lima} using the temperature-redshift 
relation derived by \cite{jetzer2011}. The constrain is given in terms
of the effective equation of state of a hypothetical decaying 
dark energy model. From the bounds
on the deviation from adiabatic invariance we find $w_{eff}=-1.005\pm 0.008$,
fully compatible with the concordance $\Lambda$CDM model. This constrain 
represents a 20\% improvement over the latest measurement of \cite{spt}.

\subsection{Possible systematic effects}

Our estimates do not take into account the effect of subtracting the
TSZ effect from the LGMCA map using the $G(\nu,\alpha=0)$ frequency dependence.
Fig.~\ref{fig3}a shows that the average temperature anisotropy at
the cluster locations has $\sim -1\mu$K residual compared with the same measurement
at random positions in the sky averaged over 100 realizations. 
This difference is well within the errors and compatible with sample variance. 
We checked that the average on the six redshift bins was 
random, oscillating around the mean, indicating that there was not such
systematic effect.  Nevertheless, if the redshift evolution has $\alpha\ne 0$ 
then there would exist a residual TSZ at the cluster location and
we need to consider the biases introduced if the $-1\mu$K difference was 
in fact systematic.  Since the LGMCA map is a combination of WMAP
and Planck data at different frequencies, when subtracting the same LGMCA map
to all Planck Nominal maps, the residual TSZ anisotropy would shift the zero 
cross frequency but will not change the dependence on $\alpha$. 
The magnitude and sign of the temperature shift is difficult to predict
since it would depend on the 
cluster redshift, mass and extent and on the LGMCA mixing matrices estimated
for a set of input channels on a patch of data at a given wavelet scale.
We reanalyzed the data subtracting a fix $1\mu$K from the measured temperature 
anisotropies of all the clusters and at all frequencies and obtained that the 
value of $\alpha$ decreased by $\delta\alpha=-0.02$ in the fit method
and remained within the errors in the ratio method; i.e.,
we would be obtaining a value closer to $\alpha=0$ by this amount. If
$\alpha>0$ then the residual TSZ at the cluster location would be
positive and so would be the bias in $\alpha$.  Therefore,
by using the TSZ frequency dependence of an adiabatically evolving Universe,
our method could be masking the effect of a non-adiabatic evolution.
Taking into account this effect, our final constraint would be
$\alpha=-0.007\pm 0.013\; (-0.02)$
where the parenthesis indicates the systematic contribution. At the
2$\sigma$ level, this result is both compatible with all the upper
limits from TSZ and spectral lines.

\section{Conclusions.}

We have constrained the deviation of the adiabatic evolution of
the CMB black-body temperature applying two different estimators
to a sample of X-ray selected clusters and using foreground cleaned 
Planck Nominal maps. By not including clusters selected by their
TSZ signature, we avoid biasing our sample to those clusters
that are closest to adiabatic evolution. Following \cite{hurier} 
we distributed our cluster in six bins of redshift. The constrains
using the ratio and fit methods were $\alpha=-0.03\pm 0.06$ and
$\alpha=-0.007\pm 0.013$, respectively; the fit method producing 
statistically more significant results than the 
ratio method since the latter is very sensitive to small denominators. 
The constrains are weakened if we add a hypothetical systematic
effect due to the component separation method used to construct
the LGMCA map and become $\alpha=-0.007\pm 0.013\; (-0.02)$
and $\alpha=-0.03\pm 0.06\; (-0.02)$, compatible with the
results given in the literature at the $2\sigma$ confidence level.

Compared with \citet{spt} the errors in the ratio method are at least a factor
of two worse. While our sample contains four times as many clusters, they are located
at much lower redshift, where the statistic is less sensitive. However,
our implementation of the fit method provides the best constraints on
$\alpha$ to date using only CMB data. For instance,
as illustrated in Fig~\ref{fig7}, by using the LGMCA reconstruction of the 
intrinsic CMB temperature anisotropies to remove the cosmological contribution 
from our maps, we have reduced the errors compared with \citet{hurier}
and obtained the best bound to date. The constrain is also interesting
since our cluster sample has the lowest mean redshift. We constrain
deviations of adiabatic evolution within $z\le 0.3$, well in the
period of accelerated expansion.
Our results are comparable with a similar analysis carried out by \citet{luzzi15}
using a catalog of clusters at higher redshifts and fully complements their findings.

Since our clusters have been selected
using X-ray data, they are not affected by possible biases associated
with selecting clusters using the TSZ effect, providing a consistency
check to previous results.
We continue to expand our X-ray cluster sample to probe higher
redshifts, where systematic biases would be larger and could be
estimated better. \cite{chluba} has argued that low redshift energy 
injections that could produce a non-adiabatic evolution of the Universe
are strongly constrained by the upper limit set by FIRAS. Nevertheless, 
and independently of theoretical expectations, measurements of the 
temperature-redshift relation of the CMB black-body temperature 
provide a strong consistency check of the current Big-Bang paradigm.

\section{Acknowledgements}

We thank the referee for his/her comments that helped to improve the 
manuscrip. This work was done in the context of the FCT/MICINN cooperation grant
'Cosmology and Fundamental Physics with the Sunyaev-Zel'dovich Effect'
AIC10-D-000443, with additional support from projects 
FIS2012-30926 from the Ministerio de Educaci\'on y 
Ciencia, Spain and PTDC/FIS/111725/2009 from FCT, Portugal.
CM is also supported by a FCT Research Professorship, contract
reference IF/00064/2012, funded by FCT/MCTES (Portugal) and POPH/FSE (EC).

\end{document}